\documentclass[10pt,aps,prb,twocolumn,noshowpacs]{revtex4-1}
\usepackage{amsmath,graphicx,amsbsy,amssymb,amsmath,epsfig,latexsym,wasysym,pifont,mathrsfs,natbib}
\usepackage{float} \newfloat{widefig}{thp}{lop} \usepackage{comment}
\usepackage{array, makecell} \usepackage{verbatim} \usepackage{datetime}
\usepackage{multirow,array} \usepackage{hyperref} \usepackage{color} \usepackage{setspace}
\usepackage{subcaption}
\usepackage{graphicx}
\usepackage{dirtytalk}
\usepackage[british]{babel}
\usepackage[utf8,latin1]{inputenc}
\UseRawInputEncoding
\usepackage{csquotes}
\usepackage{comment}
\usepackage[export]{adjustbox}

\DeclareMathOperator*{\dprime}{\prime \prime}
\hypersetup{colorlinks,
  linkcolor=blue,%
  citecolor=blue,%
  urlcolor=blue}

\begin{document}

\title{\textbf{Strain aided drastic reduction in lattice thermal conductivity and improved thermoelectric properties in Janus MXenes }}

\author{Himanshu Murari}
\email[]{hmurari@iitg.ac.in}
\affiliation{Department of Physics, Indian Institute of Technology
  Guwahati, Guwahati-781039, Assam, India.} 
\author{Swati Shaw}
\email[]{swatishaw@iitg.ac.in}
\affiliation{Department of Physics,
  Indian Institute of Technology Guwahati, Guwahati-781039, Assam,
    India.} 
\author{Subhradip Ghosh}
\email{subhra@iitg.ac.in} \affiliation{Department of Physics,
  Indian Institute of Technology Guwahati, Guwahati-781039, Assam,
    India.} 
\begin{abstract}  
 Surface and strain engineering are among the cheaper ways to modulate structure property relations in materials. Due to their compositional flexibilities, MXenes, the family of two-dimensional materials, provide enough opportunity for surface engineering. In this work, we have explored the possibility of improving thermoelectric efficiency of MXenes through these routes. The Janus MXenes obtained by modifications of the transition metal constituents and the functional groups passivating their surfaces are considered as surface engineered materials on which bi-axial strain is applied in a systematic way. We find that in the three Janus compounds Zr$_{2}$COS, ZrHfO$_{2}$ and ZrHfCOS, tensile strain modifies the electronic and lattice thermoelectric parameters such that the thermoelectric efficiency can be maximised. A remarkable reduction in the lattice thermal conductivity due to increased anharmonicity and elevation in Seebeck coefficient are obtained by application of moderate tensile strain. With the help of first-principles electronic structure method and semi-classical Boltzmann transport theory we analyse the interplay of structural parameters, electronic and dynamical properties to understand the effects of strain and surface modifications on thermoelectric properties of these systems. Our detailed calculations and in depth analysis lead not only to  the microscopic understanding of the influences of surface and strain engineering in these three systems, but also provide enough insights for adopting this approach and improve thermoelectric efficiencies in similar systems. 
 \end{abstract}


\maketitle

\section{Introduction\label{intro}}
In the era of sustainable energy, thermoelectric (TE) material is one of the prominent candidate serving the purpose, thus inviting research focus in recent years\cite{sootsman2009new,mori}.  Usage of TE devices on a large scale is however hindered due to their limited conversion efficiency of heat into electricity. The efficiency of a TE material is assessed by a dimensionless quantity known as the figure of merit  $ZT = \frac{S^{2}\sigma}{\kappa_{e}+\kappa_{l}}T$, where $S$ is the Seebeck coefficient, $\sigma$ the electrical conductivity, $\kappa_{e}$ and $\kappa_{l}$ the electronic and lattice thermal conductivity, respectively, and $T$ the operational temperature\cite{disalvo1999thermoelectric}. Due to the strong interdependence of the transport coefficients, enhancing $ZT$ is challenging; the optimization of one coefficient directly affects the other as $\sigma$ is directly and inversely related to $S$ (Mott's relation\cite{motts}) and $\kappa_{e}$ (Wiedemann-Franz law\cite{Wiedemann}), respectively. The only quantity independent of other transport coefficients is $\kappa_{l}$. Consequently, a possible way to enhance ZT can be achieved by reducing it\cite{yu2019ultralow}. Much effort has been made to increase the $ZT$ through band convergence\cite{pei2011convergence} and nanostructuring\cite{biswas2012high,thinfilmte}. In this context, low-dimensional materials, especially two-dimensional (2D) materials, have attracted great attention\cite{hicks1993effect,hicks1996experimental}. The quantum confinement effects in 2D materials affect the transport coefficients: large densities of states (DOS) near the Fermi level causes significant increase in $S$\cite{dresselhaus2007new}, and enhanced phonon-phonon scattering reduces $\kappa_{l}$\cite{alam2013review}, providing a possible solution to the challenge of increasing efficiencies in TE materials. Subsequently, thermoelectric properties of a number of 2D materials like Graphene, transition metal dichalcogenides, h-BP and group IV-VIA layered compounds  were extensively explored \cite{tmdc,hbp,snse}. The results were encouraging.

Relative new addition to the list of 2D materials is MXene, the family of transition metal carbonitrides. MXenes are the 2D layered counterparts of three-dimensional MAX compounds, obtained by removing the A layers inter-connecting the transition metal M and carbon/nitrogen X layers in MAX, through exfoliation. During this process, various functional groups passivate the dangling bonds on the surfaces rendering MXenes with a chemical formula of M$_{n+1}$X$_{n}$T$_{x}$ (n=1-3), where M, X, and T are the transition metal, carbon/nitrogen, and surface functional groups, respectively. Naturally, as compared to other 2D materials, MXenes offer a greater compositional flexibility and tunability in the functional properties. Such a prospect, with regard to thermoelectric properties, has been explored aggressively for a number of MXenes, both experimentally and with first-principles simulations\cite{filmti3c2,sc2c,gandi2016thermoelectric,sarikurt2018influence,tanusri,jing2019superior}. It was found that the compositional and structural tuning brought about by changes in the passivating sites, the passivating functional groups or the compositions at the transition metal sites lower the lattice thermal conductivity $\kappa_{l}$ considerably, increasing $ZT$ as a consequence. The lowering of symmetry due to such tuning led to increased phonon-phonon scattering, and increased anharmonicity resulting in lowering of $\kappa_{l}$. The effects of chemical substitution and lowering of symmetry on TE properties were also observed in experimentally synthesised Janus 2D TMDC MoSSe \cite{lu2017janus,deng2019enhanced}. The asymmetry in the two surfaces of MoX$_{2}$ (X=S,Se) introduced by different functional groups passivating different surfaces led to the out-of-plane-ordered Janus structure  and gave rise to superior TE properties in comparison with the end-point TMDCs MoX$_{2}$. An enhanced in the $ZT$ was also found in Janus-PdSeTe compared to its end point compounds\cite{pdsete1,pdsete2}. In the MXene family, high $ZT (\sim 3$) was obtained for Janus compounds TiZrCO$_{2}$ and TiHfCO$_{2}$\cite{murari2024symmetry}, suggesting that such out-of-plane ordering in MXenes can be explored to achieve thermoelectric materials with superior properties. 

Strain engineering too is found to be a low-cost and feasible method to enhance the thermoelectric properties of 2D materials \cite{strain1,strain2,strain3,strain4}. The effect of strain is prominently seen in the electronic band structures and phonon transport. In 1T-TiS$_{2}$ monolayer, semi-metal to semiconductor transition is observed under tensile strain . In this compound, a 6\% tensile strain opens up a band gap of 0.57 eV\cite{tis2}. Band convergence can be achieved by application of strain that in turn affects the electronic transport coefficients. Large thermopower ($S$) ranging from 250-380 and 390-440 $\mu$V/K are reported for few-layered MoS$_{2}$\cite{strain4}; the reason attributed to tunable band degeneracy under different types of strain. Uniaxial strain-induced band convergence for phosphorene monolayer has been reported as an effective method to enhance $ZT$ to 2.12 at 300 K\cite{phosphorene}. It is also found that the tensile biaxial strain reduces  $\kappa_{l}$ by nearly $50 \%$ \cite{strainreview,ptse2}. Due to increased phonon-phonon scattering, it was found that the tensile strain reduces the $\kappa_{l}$ from 16.97 W/m-K to 6.88 W/m-K at 300 K for PtSe$_{2}$\cite{ptse2}.  A 7\% tensile strain in p-type HfS$_{2}$ was found to lead to a maximum $ZT$ of approximately 3.3 \cite{hfs2}.

Motivated by these findings, in this work, we have investigated the effects of structural modifications and strain on the thermoelectric properties of MXenes. The structural modification is considered by constructing three different Janus compositions in Zr and Hf based M$_{2}$CT$_{2}$ MXenes: (a) Zr$_{2}$COS Janus, a derivative of Zr$_{2}$CO$_{2}$ where one of the functional group -O is replaced with -S, (b) ZrHfCO$_{2}$ Janus, where Zr on one surface is completely replaced with Hf and (c) ZrHfCOS Janus where along with replacement of one of the transition metal surface in Zr$_{2}$CO$_{2}$, one functional group is also replaced that is the Janus is made with respect to both M and T of M$_{2}$CT$_{2}$. The reason behind choosing Zr$_{2}$CO$_{2}$ as the base compound is the reported high (moderate) value of $S$ ($\kappa_{l}$) in this compound \cite{gandi2016thermoelectric,murari2024symmetry}. The thermoelectric parameters of these three compounds as a function of bi-axial strain are calculated by first-principles Density Functional Theory \cite{hohenberg1964inhomogeneous}in conjunction with Boltzmann transport theory. The thermoelectric properties are explored at different carrier concentrations  and  different temperatures (300-800 K). We find that in all three compounds effects of surface manipulation and tensile strain manifest in lowering of $\kappa_{l}$ and elevation of $S$ by considerable amount. A maximum $ZT$ of 3.2 is found for n-type ZrHfCO$_{2}$ at 800 K. For the other two compounds too, strain leads to maximum $ZT \sim 2$. We have presented a systematic in-depth analysis of the thermoelectric parameters to explain the remarkable gain in them for these MXenes.

\section{Methodology\label{methods}}

First-principle calculations have been performed within the density functional theory (DFT)\cite{hohenberg1964inhomogeneous,kohn1965self} based Projector Augmented Wave method\cite{blochl1994projector} implemented in the Vienna Ab-initio Simulation Package (VASP)\cite{kresse1996efficient}. The exchange-correlation part of the Hamiltonian is described by the Perdew-Burke-Ernzerhof (PBE)\cite{perdew1996generalized} Generalized Gradient Approximation (GGA). During the structural optimization, the cutoff energy for plane wave basis is chosen to be 550 eV. For the Brillouin zone sampling, a 16$\times$16$\times$1 $\Gamma$-centered k-mesh is employed\cite{monkhorst1976special}. The convergence criteria for the forces/atom and total energy are taken to be 10$^{-3}$ eV/{\AA} and 10$^{-7}$ eV. A denser k-mesh of 32$\times$32$\times$1 is chosen for densities of states calculation. A vacuum of 20 {\AA}  is considered along the $c$-axis to avoid  interactions between periodic images. Ab initio molecular dynamics (AIMD) calculations are performed to ascertain thermodynamical stability of the compounds. The AIMD simulations are carried out using a canonical ensemble (NVT) in the Nose Hoover heat bath\cite{nose1984unified}. The simulations are performed on a 3$\times$3$\times$1 supercell with a 4$\times$4$\times$1 $k$-point grid. Simulations are run for 12 ps.

The electronic transport coefficients (Seebeck coefficient ($S$), electrical conductivity ($\sigma$/$\tau$), and electronic thermal conductivity ($\kappa_{e}$/$\tau$)) are evaluated by solving the  Boltzmann transport equations (BTE) under constant relaxation time approximation (CRTA) and rigid band approximation (RBA), as implemented in the BoltzTrap2 package\cite{madsen2018boltztrap2}. In the CRTA approach, the electronic relaxation time ($\tau$) is held constant while evaluating the transport coefficients. In the RBA, the material's electronic structure only undergoes a rigid shift when temperature or carrier concentration changes. The expressions for electrical conductivity tensor ($\sigma_{\alpha\beta}(T,\mu)$), electronic thermal conductivity tensor ($\kappa^{o}_{\alpha\beta}(T,\mu)$), and Seebeck coefficient tensor ($S_{\alpha\beta}(T,\mu)$) are given as
\begin{equation}\label{sig}
	\sigma_{\alpha\beta}(T,\mu)= \frac{1}{\Omega}\int {\bar{\sigma}_{\alpha\beta}(\epsilon)}\Big{[}-{\frac{\partial f(T,\epsilon,\mu)}{\partial\epsilon}}\Big{]}d\epsilon
\end{equation},
\begin{equation}\label{kappae}
	\kappa^{o}_{\alpha\beta}(T,\mu)=\frac{1}{e^{2}T\Omega}\int{\bar{\sigma}_{\alpha\beta}(\epsilon)}(\epsilon - \mu)^{2}\Big{[}-{\frac{\partial f(T,\epsilon,\mu)}{\partial\epsilon}}\Big{]}d\epsilon
\end{equation}

\begin{equation}\label{seebeck}
	S_{\alpha\beta}(T,\mu)=\frac{1}{eT\Omega\sigma_{\alpha\beta}(T,\mu)}\int{\bar{\sigma}_{\alpha\beta}(\epsilon)}(\epsilon-\mu)\Big{[}-{\frac{\partial f(T,\epsilon,\mu)}{\partial\epsilon}}\Big{]}d\epsilon
\end{equation}
where
\begin{equation}\label{bar_sigma}
	{\bar{\sigma}}_{\alpha\beta}(\epsilon) = \frac{e^{2}}{N}\sum_{i,k}{\tau v_{\alpha}(i,k) v_{\beta}(i,k)\delta(\epsilon-\epsilon_{i,k})}
\end{equation}
and
\begin{equation}\label{v}
	v_{\alpha}(i,k)=\frac{1}{\hbar}\frac{\partial\epsilon_{i,k}}{\partial k_{\alpha}}
\end{equation}

 $\Omega$ is the volume of the unit cell, $\mu$ is the chemical potential, and $f$ is the Fermi-Dirac distribution function. $N$ denotes the number of $k$ points sampled, and $e$  the electron charge. $v_{\alpha}(i,k)$($\alpha=x,y,z$) indicates the $\alpha$-th component of the group velocity of the carriers, corresponding to the $i$-th energy band. Equations (\ref{sig})-(\ref{v}) suggest that the transport coefficients can be evaluated using the group velocity $v_{\alpha}(i,k)$ obtained from the band structure computed with DFT.

The second-order (harmonic) interatomic force constants (IFCs) and phonon dispersions are obtained by the supercell approach in conjunction with the finite displacement method as implemented in the Phonopy package\cite{togo2015first}. A well converged  6$\times$6$\times$1 supercell with 4$\times$4$\times$1 $k$-point grid has been used. The lattice thermal conductivities have been calculated by solving the phonon Boltzmann transport equation, using an iterative scheme implemented in ShengBTE code\cite{li2014shengbte}. The lattice thermal conductivity is given by,
\begin{equation}
    \kappa^{\alpha\beta}_{l} = \frac{1}{k_{B}T^{2}VN}\sum_{\lambda}f_{0}(f_{0}+1)(\hbar\omega_{\lambda})^{2}v_{\lambda}^{\alpha}F_{\lambda}^{\beta}
    \label{kl_eq}
\end{equation}
 $V$ is the volume of the unit cell, and $N$ the number of $q$ points sampled in the Brillouin zone.  $\omega_{\lambda}$ and $v_{\lambda}^{\alpha}$ are the angular frequency and group velocity of the phonon branch $\lambda$ along the $\alpha$ direction, respectively. $f_{0}$ is the Bose distribution function, and $F_{\lambda}^{\beta}=\tau_{\lambda}^{0}(v_{\lambda}^{\beta}+\Delta_{\lambda}^{\beta})$ is the form of linearised BTE when only two and three phonon scatterings are considered. $\Delta_{\lambda}^{\beta}$ is the correction term obtained by a fully iterative solution of the BTE. $v_{\lambda}^{\beta}$ is group velocity of the phonon branch $\lambda$ along the $\beta$ direction. $\tau_{\lambda}^{0}$ is the relaxation time of phonon mode $\lambda$, the inverse of which is equal to the sum of all possible transition probabilities between phonon modes $\lambda$, $\lambda^{\prime}$ and $\lambda^{\dprime}$. 
\begin{equation}
    \frac{1}{\tau_{\lambda}^{0}} = \frac{1}{N}\Big{(}\sum_{\lambda^{\prime}\lambda^{\dprime}}^{+}\Gamma_{\lambda\lambda^{\prime}\lambda^{\dprime}}^{+} + \sum_{\lambda^{\prime}\lambda^{\dprime}}^{-}\frac{1}{2}\Gamma_{\lambda\lambda^{\prime}\lambda^{\dprime}}^{-} + \sum_{\lambda^{\prime}}\Gamma_{\lambda\lambda^{\prime}}\Big{)}
\end{equation}
The first two terms in the above equation correspond to  three-phonon scattering rates. The first one describes absorption and the second term  the emission processes. The third term corresponds to isotopic scattering. The expressions for absorption (+) and emission (-) processes are given as
\begin{equation}
    \Gamma_{\lambda\lambda^{\prime}\lambda^{\dprime}}^{+} = \frac{\hbar\pi}{4}\frac{f_{0}^{\prime}-f_{0}^{\dprime}}{\omega_{\lambda}\omega_{\lambda^{\prime}}\omega_{\lambda^{\dprime}}}|\phi_{\lambda\lambda^{\prime}\lambda^{\dprime}}^{+}|^{2}\Delta(\omega_{\lambda}+\omega_{\lambda^{\prime}}-\omega_{\lambda^{\dprime}})
\end{equation}
\begin{equation}
    \Gamma_{\lambda\lambda^{\prime}\lambda^{\dprime}}^{-} = \frac{\hbar\pi}{4}\frac{f_{0}^{\prime}+f_{0}^{\dprime}+1}{\omega_{\lambda}\omega_{\lambda^{\prime}}\omega_{\lambda^{\dprime}}}|\phi_{\lambda\lambda^{\prime}\lambda^{\dprime}}^{-}|^{2}\Delta(\omega_{\lambda}-\omega_{\lambda^{\prime}}-\omega_{\lambda^{\dprime}})
\end{equation}
 $|\phi_{\lambda\lambda^{\prime}\lambda^{\dprime}}^{\pm}|$ are the scattering matrix elements that depend on the anharmonic IFCs ($\Phi_{ijk}^{\alpha\beta\gamma}$) as,
\begin{equation}
    |\phi_{\lambda\lambda^{\prime}\lambda^{\dprime}}^{\pm}| = \sum_{i\epsilon u.c.} \sum_{j,k} \sum_{\alpha\beta\gamma} \Phi_{ijk}^{\alpha\beta\gamma}\frac{e_{\lambda}^{\alpha}(i) e_{\lambda^{\prime}}^{\beta}(j) e_{\lambda^{\dprime}}^{\gamma}(k)}{\sqrt{M_{i}M_{j}M_{k}}}
    \label{eq5}
\end{equation}
 $M_{i}$ is the mass of $i^{th}$ atom,  $e_{\lambda}^{\alpha}(i)$ denotes the $\alpha$-th component of an eigenvector of mode $\lambda$ associated with the $i$-th atom. Here summation over $i$ is within the unit cell, while summation over $j$ and $k$ span the entire system.

The third-order IFCs are evaluated using a supercell of 4$\times$4$\times$1 with eight nearest neighbors. The harmonic and anharmonic IFCs  obtained are used to compute the lattice thermal conductivity $\kappa_{l}$ and solve BTE for phonons. To evaluate accurate $\kappa_{l}$, the harmonic IFCs are corrected by enforcing the rotational sum rule, implemented in machine learning-based hiphive package \cite{hiphive}. For calculating $\kappa_{l}$, a denser $q$-grid of 100$\times$100$\times$1 is used. The $\kappa_{l}$ thus obtained is scaled by a factor $c/z$, where $c$ is vacuum height and $z$  the thickness of the material considered. The convergence of $\kappa_{l}$ with respect to sizes of the supercell and the $q$-grid is checked by separate calculations on supercells of size 3$\times$3$\times$1 and 4$\times$4$\times$1, $q$-grids of size 90$\times$90$\times$1 and 100$\times$100$\times$1. The differences are found to be about 1\% only.

\section{Results and discussion}
\subsection{Structural parameters and stability}
\begin{figure}
    \centering
    \includegraphics[scale=0.3]{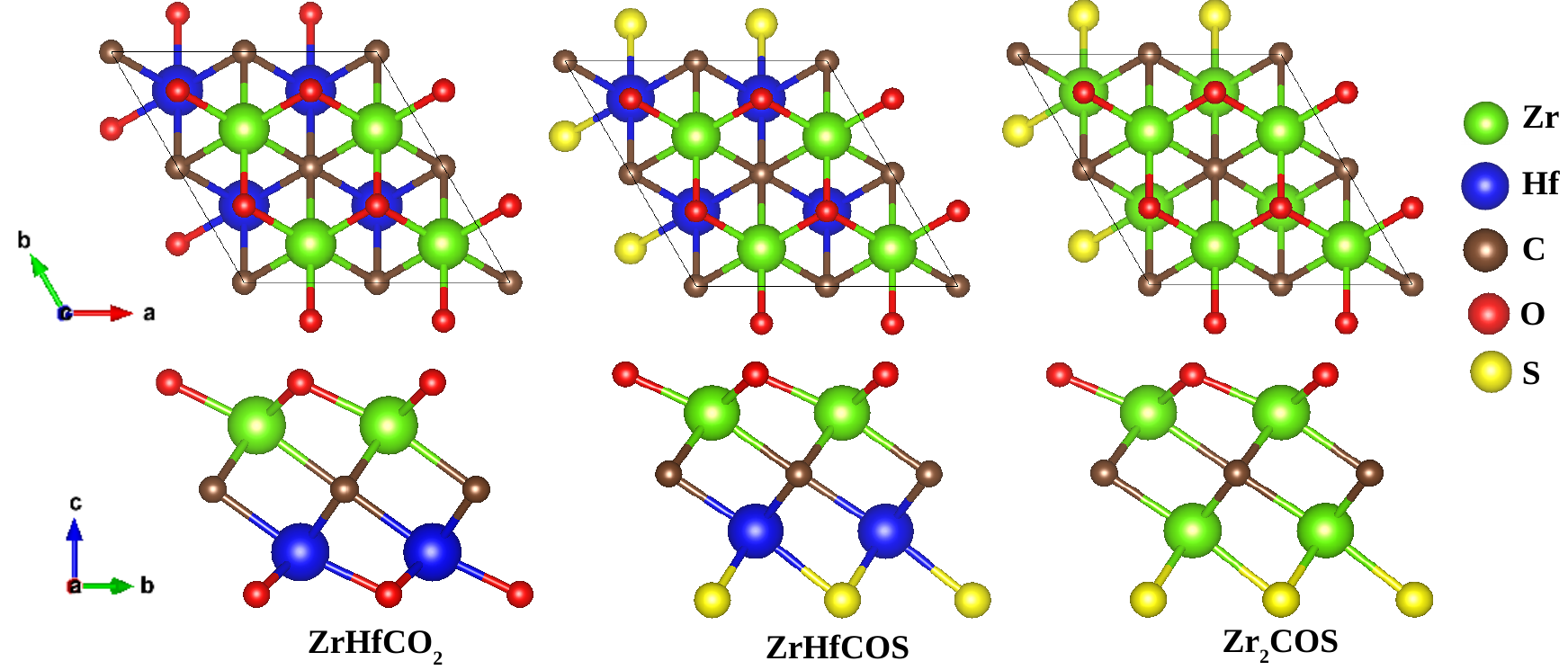}
    \caption{Ground state structures of Janus MXenes. Top (bottom) panel shows the top (side) view}
    \label{fig1}
\end{figure}

The Janus monolayers considered here are obtained from Zr$_{2}$C pristine MXene. Figure S1 shows the possible structural models with all surface sites available to -O and -S atoms (details are discussed in supplementary information). The ground state structures are obtained from DFT calculations by calculating total energies corresponding to all possible structural models. The structural model with minimum total energy is considered as the ground state. We find that both -O and -S atoms occupying the H site (the hollow site corresponding to the transition metal atoms) minimise the total energy in all cases. The ground state structure in each of the three Janus compounds is shown in Figure \ref{fig1}.
\vspace{0.5mm}
 \begin{table}
	\centering
	\resizebox{1.0\columnwidth}{!}{%
		\begin{tabular}{|c|c|c| c| c|c| c| c|}
            
            \hline
            
            \textbf{Compound}  & \textbf{Strain}  & $\textbf{a}$ & \textbf{E$_{g}$} & \multicolumn{4}{c|}{\multirow{2}{*}{\textbf{Bond length (\AA)}}} \\
                      & \textbf{($\epsilon$)} \% & (\AA) & (eV) & \multicolumn{4}{c|}{} \\
            \hline
            \hline
            \multirow{6}{*}{ZrHfCO$_{2}$}& & & & Zr-C & Hf-C & Zr-O$_{u}$ & Hf-O$_{d}$ \\
            \cline{2-8}
            & -4 &3.16 &0.61 &2.30 &2.28 &2.04 &2.03 \\
            & -2 &3.22 &0.82 &2.33 &2.31 &2.07 &2.07 \\
            &  0 &3.29 &0.99 &2.36 &2.34 &2.11 &2.10 \\
            & +2 &3.35 &1.13 &2.39 &2.37 &2.14 &2.14 \\
            & +4 &3.42 &1.24 &2.42 &2.40 &2.18 &2.17 \\
            \hline
            \hline
            \multirow{6}{*}{ZrHfCOS}& & & & Zr-C & Hf-C & Zr-O & Hf-S \\
            \cline{2-8}
            & -4 &3.23 &- &2.35 &2.28 &2.10 &2.49 \\
            & -2 &3.30 &- &2.37 &2.31 &2.11 &2.49 \\
            &  0 &3.37 &- &2.40 &2.34 &2.13 &2.50 \\
            & +2 &3.44 &0.02 &2.43 &2.37 &2.16 &2.51 \\
            & +4 &3.50 &0.22 &2.45 &2.40 &2.18 &2.52 \\
            \hline
            \hline 
            \multirow{6}{*}{Zr$_{2}$COS}& & & & Zr$_{u}$-C & Zr$_{d}$-C & Zr$_{u}$-O & Zr$_{d}$-S \\
            \cline{2-8}
            & -4 &3.25 &- &2.35 &2.31 &2.10 &2.52 \\
            & -2 &3.32 &- &2.38 &2.34 &2.12 &2.52 \\
            &  0 &3.39 (3.44\cite{WANG}) &- &2.40 &2.36 &2.14 &2.52 \\
            & +2 &3.46 &- &2.43 &2.39 &2.16 &2.53 \\
            & +4 &3.52 &0.18 &2.46 &2.46 &2.18 &2.54 \\
            \hline
		\end{tabular}
  }	
        \begin{minipage}{8.7cm}
            \footnotesize{$^{*}$Zr$_{d}$ (O$_{d}$) and Zr$_{u}$ (O$_{u}$) are the Zr (O) atoms at the (00-1) and (001) surfaces of Zr$_{2}$COS (ZrHfCO$_{2}$) MXene, respectively}
        \end{minipage}
        \caption{ Lattice constant (a), band gap (E$_{g}$), and bond lengths corresponding to different pairs of atoms in the three Janus MXenes.Results are tabulated for different $\epsilon$.}
        \label{tab1}
\end{table}

In Table \ref{tab1}, the structural parameters of the systems as a function of bi-axial strain $\epsilon (\epsilon=(\frac{a-a_{0}}{a_{0}})\times100\%$), where $a_{0}$(a) is the lattice constant for unstrained (strained) system) are shown. To ensure stability of the compounds, the range of strain considered here is kept limited between -4\% and +4\%. We find significant dispersions in the M-C and M-T bond lengths. When the functional group T is -O (-S), the M-C bonds are  about 13\% (2-7\%) longer (shorter) than the M-T bonds.  Such fluctuations in bond lengths as an effect of surface manipulation introduce anharmonicity in the system and directly affect the lattice thermal conductivity. 
\begin{table}
		\centering
		\begin{tabular}{|c|c|c|c|c|}
			\hline
			 & \textbf{Strain} & \multicolumn{3}{c}{\textbf{Elastic Constants}}\vline\\
			\cline{3-5}
			 \textbf{System}& \textbf{($\epsilon$)} & \textbf{C$_{11}$=C$_{22}$}& \textbf{C$_{12}$} & \textbf{C$_{66}$} \\
			&(\%) & (Nm$^{-1}$) & (Nm$^{-1}$)&(Nm$^{-1}$) \\
			\hline
			\hline
			\multirow{5}{*}{ZrHfCO\textsubscript{2}}&-4&339.16&95.95&121.61\\
			\cline{2-5}
			&-2&311.06&85.82&112.62\\
			\cline{2-5}
			&0&278.19&78.51&99.84\\
			\cline{2-5}
			&+2&246.94&72.04&87.45\\
			\cline{2-5}
			&+4&214.34&67.60&73.37\\
			\hline
			\hline
			\multirow{5}{*}{ZrHfCOS}&-4&256.38&68.36&94.01\\
			\cline{2-5}
			&-2&232.63&63.69&84.47\\
			\cline{2-5}
			&0&219.35&61.55&78.90\\
			\cline{2-5}
			&+2&196.14&56.86&69.64\\
			\cline{2-5}
			&+4&170.05&57.67&56.19\\
			\hline
			\hline
			\multirow{5}{*}{Zr\textsubscript{2}COS}&-4&252.81&72.46&90.18\\
			\cline{2-5}
			&-2&210.54&56.54&77.00\\
			\cline{2-5}
			&0&213.66&63.42&75.12\\
			\cline{2-5}
			&+2&185.22&56.05&64.59\\
			\cline{2-5}
			&+4&163.97&56.36&53.80\\
			\hline
		\end{tabular}
        \caption{{The Calculated Elastic Constants (\textbf{C$_{11}$}, \textbf{C$_{22}$}, \textbf{C$_{12}$}, \textbf{C$_{66}$}) as a function of strain are shown for the three compounds.}}
	\label{tab2}
\end{table}

Since the Janus MXenes considered in this work are yet to be synthesised experimentally,their mechanical, thermal and dynamical stabilities as a function of strain are crucial. In what follows,  we have investigated these by varying the strain from compressive to the tensile region. The mechanical stability of the optimized structures at each strain is examined by computing the elastic constants C$_{ij}$. Since the symmetry of the Janus structure is hexagonal, three  elastic constants  C$_{11}$, C$_{12}$ and  C$_{66}=$(C$_{11}$-C$_{12}$)/2 are computed. Table \ref{tab2} shows the evaluated elastic constants of the three compounds at each value of $\epsilon$. We find that the  criteria\cite{Born_1940} for mechanical stability, C$_{11}>$ 0 and C$_{11}>$ C$_{12}$, are satisfied in each case suggesting the mechanical stability of each compound for the entire range of strain considered.

\begin{figure}
    \centering
    \includegraphics[scale=0.125]{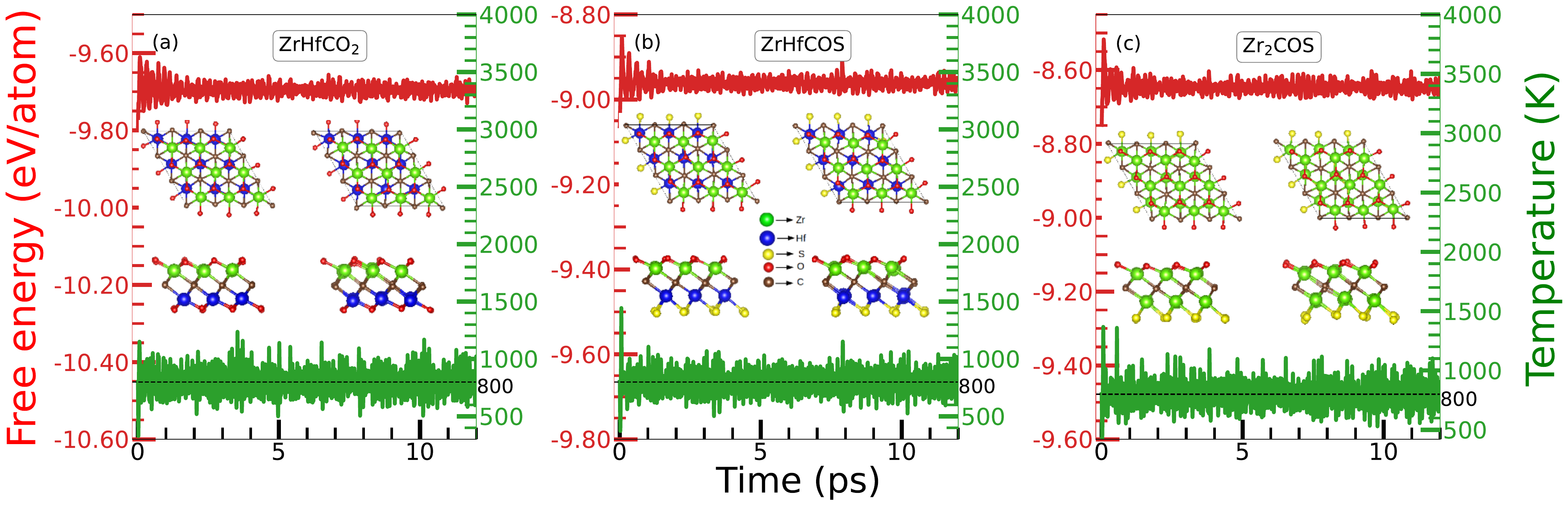}
    \caption{Variations in the Free Energy and the Temperature of the three compounds, over a time of 12 ps. The calculations are done on the systems at zero strain. The temperature considered for the calculations is 800 K.}
    \label{fig2}
\end{figure}

\begin{figure}
	\includegraphics[scale=0.31]{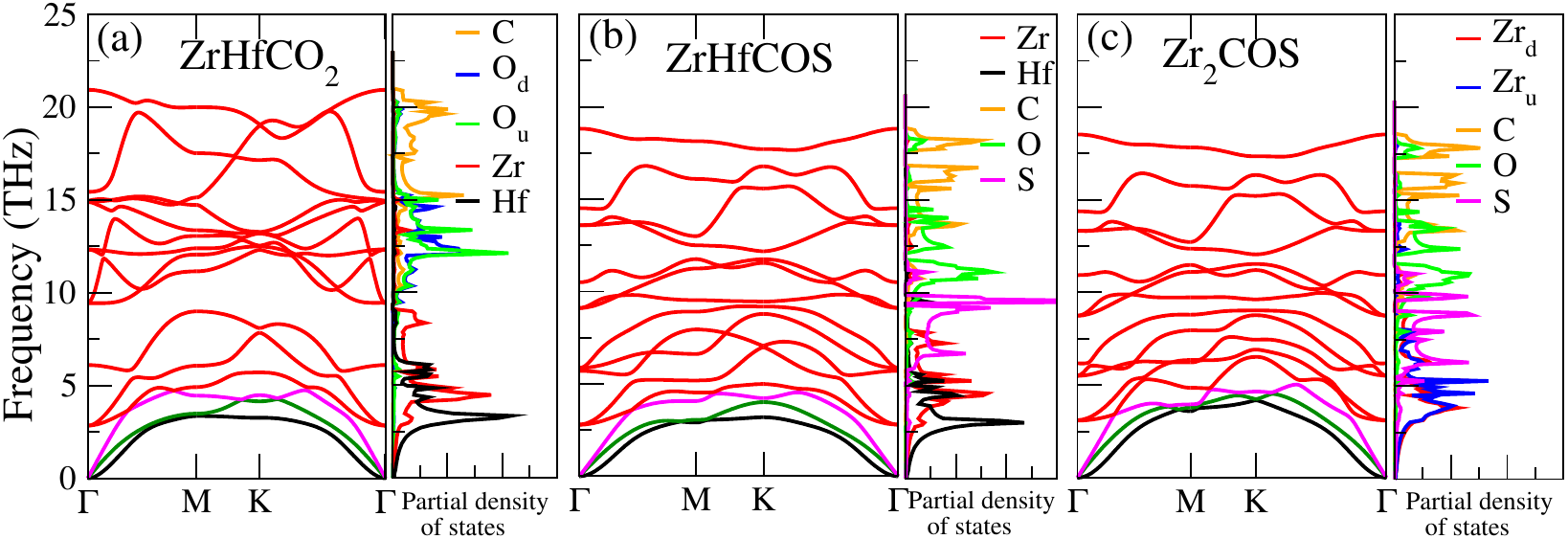}
	\caption{Phonon dispersions and atom-projected phonon densities of states for the three Janus compounds at zero strain.}
	\label{fig3}
\end{figure}

The thermal stability of a compound at higher temperatures is vital for thermoelectric (TE) applications. The thermal stability of all three Janus compounds at each strain is calculated by \textit{ab initio} molecular dynamics (AIMD) simulations. Variations in the free energy and the temperature with time at 800 K, along with the  the initial (T $=$ 0 K) and final (T $=$ 800 K) structures, are shown in Figure \ref{fig2} (for zero strain) and figure S2, supplementary information (for different strains). The results suggest that even at this high temperature, none of the structures distort over time ensuring thermal stability of all compounds, irrespective of the amount of strain. The stability at 800 K, the highest  temperature for calculations of TE parameters, also ensure that the structures are stable at lower temperatures and under various strains. The dynamical stability of the compounds on application of strains is assessed by computing the phonon dispersion relations. The results are shown in Figure \ref{fig3} (Figure S3, supplementary information) for the unstrained (strained) Janus MXenes. The results imply that all three systems with and without strains are dynamically stable. Therefore, these three systems satisfy the necessary stability criteria in the temperature window of thermoelectric operation for the range of strains used in this work.

\section{Electronic structure}
\begin{figure}
    \centering
    \includegraphics[scale=0.11]{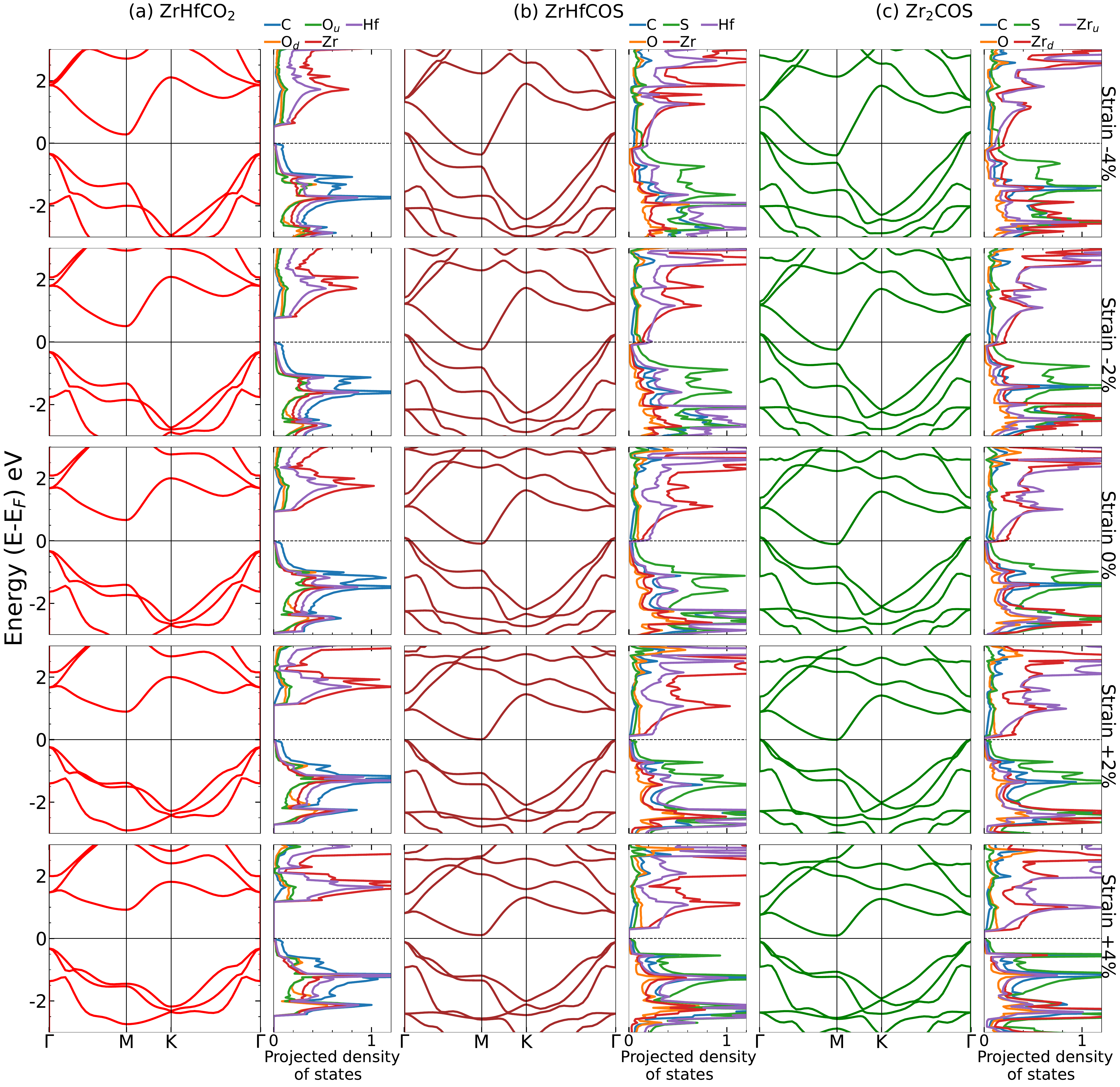}
    \caption{Electronic band structures and atom projected densities of states for the three Janus MXenes . The three panels represent (a) ZrHfCO$_{2}$, (b) ZrHfCOS, and (c) Zr$_{2}$COS, for strains  between -4\% and +4\%. }
    \label{fig4}
\end{figure}

Features in the electronic structures provide clues to the expected behaviours of electronic transport parameters associated with the thermoelectric figure of merit. In Figure \ref{fig4} we present the  electronic band structures and atom-projected densities of states of the three compounds at different strains. Without strain, ZrHfCO$_{2}$ is an indirect semiconductor whereas ZrHfCOS and Zr$_{2}$COS exhibit semi-metallic behaviour. Application of strain is quite significant for these three as the band gap changes upto about 39\% for the semiconductor and the semiconducting gaps open in the other two. The band gaps increases (decreases) with tensile (compressive) strains for all three, the largest band gap is found for a strain of +4\% strain. The evolution of the band gap with strain is shown in Table \ref{tab1}. The positions of conduction band minima (CBM) and valence band maxima (VBM) of the three compounds remain invariant with strain, the CBM (VBM) being located at the M ($\Gamma$) point. With the application of tensile (compressive) strain, the conduction band edges for all three compounds become more flat (dispersive). Moreover, the CBM for all the compounds are flatter than the VBM. This may result in a higher $S$ for n-type doping. 

The atom projected densities of states for all three compounds indicate significant contributions from $p$ orbitals of C and T and $d$ orbitals of M in both valence and conduction bands. However, the features and contributions near the band edges vary with composition and strain. For Zr$_{2}$COS and ZrHfCOS, sharp peak dominated by S shapes up with the increase in the tensile strain. Although no such sharp peak is observed in case of ZrHfCO$_{2}$, the sharpness of the band edges are found to increase with tensile strain as well. This is consistent with the flatter CBM as tensile strain is increased. This may result in larger Seebeck coefficient with tensile stress. In the next sub-section we will examine this. 

\section{Electronic transport properties}
\begin{figure}
	\includegraphics[scale=0.3]{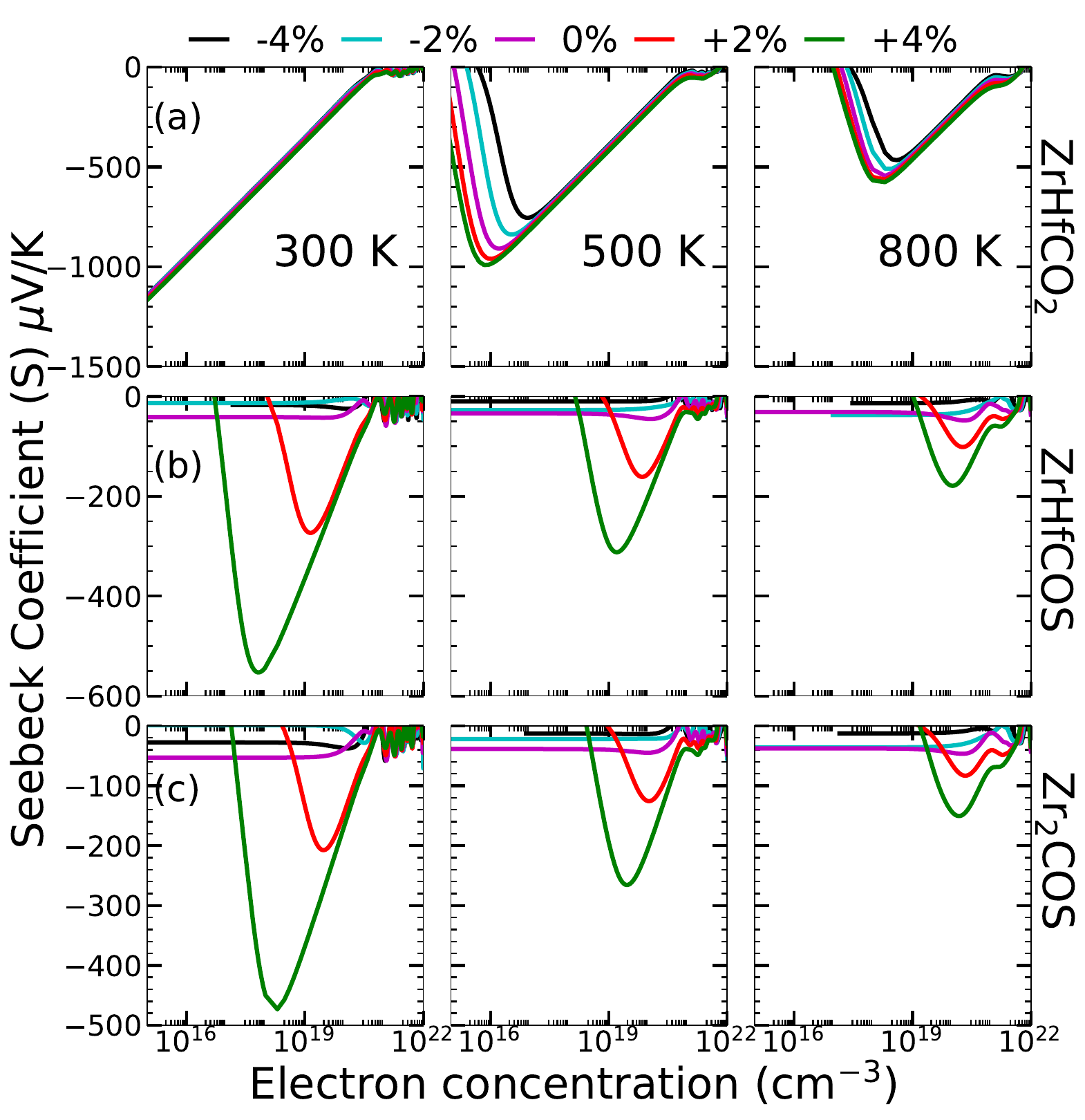}
	\caption{Seebeck coefficient (S (in $\mu$V/K)) as a function of electron concentration (in cm$^{-3}$) for n-type Janus MXenes  ((a) ZrHfCO$_{2}$, (b) ZrHfCOS, and (c) Zr$_{2}$COS) at different temperatures. Results are shown for various $\epsilon$.}
	\label{fig5}
\end{figure}

Seebeck coefficient ($S$) is the measure of the ability of a material to convert a temperature gradient into electrical voltage\cite{goldsmid}. A high value of $S$, therefore, is desired for better TE conversion efficiency. For a semiconductor, negative (positive) value of $S$ depicts that the majority-charge carriers are electrons (holes). Figure \ref{fig5} (Figure S4, supplementary information) shows the Seebeck coefficient $S$ as a function of n-type (p-type) carrier concentration at different temperatures and strains. Among the three Janus compounds, n-ZrHfCO$_{2}$ has the maximum value of $S$($\sim$ 1200 $\mu$V/K at 300 K), irrespective of strain. The qualitative variation of S with electron concentration changes as the temperature is increased. For example, at 300 K, the absolute value of $S$ of ZrHfCO$_{2}$ decreases monotonically (following the inverse relation of $S$ and $n$, the carrier concentration\cite{motts}).For higher temperatures, the variation is non-monotonic\cite{hong2016full}, that is, $S$ first increases then decreases with $n$. The position of the peak shifts with the increase in temperature. Similar behaviour is visible in ZrHfCOS and Zr$_{2}$COS even at 300 K. This behaviour of $S$ at low $n$ and high T for the narrow band gap semiconductors can be attributed to the bipolar conduction effect\cite{bipolar1}. For n-type Janus structures, although electrons are the majority carriers, the contribution from holes (minority carriers) is non-negligible. Therefore, for the bipolar conduction regime, we observe, for our narrow band gap systems, a deviation from standard $S$ $\propto$ $n^{-2/3}$ behaviour. Such deviation of $S$ due to minority carriers has been reported in earlier studies \cite{bipolar2, bipolar3, bipolar4}. 

The effect of strain is visible on all three compounds. As the tensile (compressive) strain is applied, the $S$ value increases (decreases). For n-type ZrHfCOS and n-type Zr$_{2}$COS, the effect of strain is more prominent. For these two compounds, $S$ at 4\% tensile strain is 6-8 times higher than $S$ when no strain is applied. The explanation for such enhancement in $S$ can be done from the changes in the densities of states and the band gaps (E$_{g}$). The emergence of sharp edges with application of tensile strain (Figure \ref{fig3}) and flattening of the bands increases the effective mass (m$^{*}$) leading to an increase in $S$, since $S$ $\propto$ m$^{*}$\cite{motts}, $m^{*}$ the effective mass. Such substantial effect is not seen in ZrHfCO$_{2}$ as the changes in the band edges with strain are not so drastic. Neverthelss, at any value of $\epsilon$, S$_{ZrHfCO_{2}}$ $>$ S$_{ZrHfCOS}$ $\sim$ S$_{Zr_{2}COS}$. This is due to significantly larger band gap of ZrHfCO$_{2}$ in comparison with the other two. The variations of $S$ with concentration of p-type carrier, temperature and strain are shown in Figure S4, supplementary information. The $S$ of all three compounds, obtained with p-type carriers are slightly smaller than the corresponding n-type counterpart, irrespective of the strain and temperature. This is because of the dispersive nature of valence band edges in all three compounds as compared to the corresponding conduction band edges. It is worth mentioning that  the evaluated values of $S$ for these three compounds are, in general, higher or comparable to those of some established thermoelectric materials\cite{yu2019ultralow,shu2023high,pts2}. This raises the prospect of these materials as ones with higher TE efficiency.

\begin{figure}
    \centering
    \includegraphics[scale=0.3]{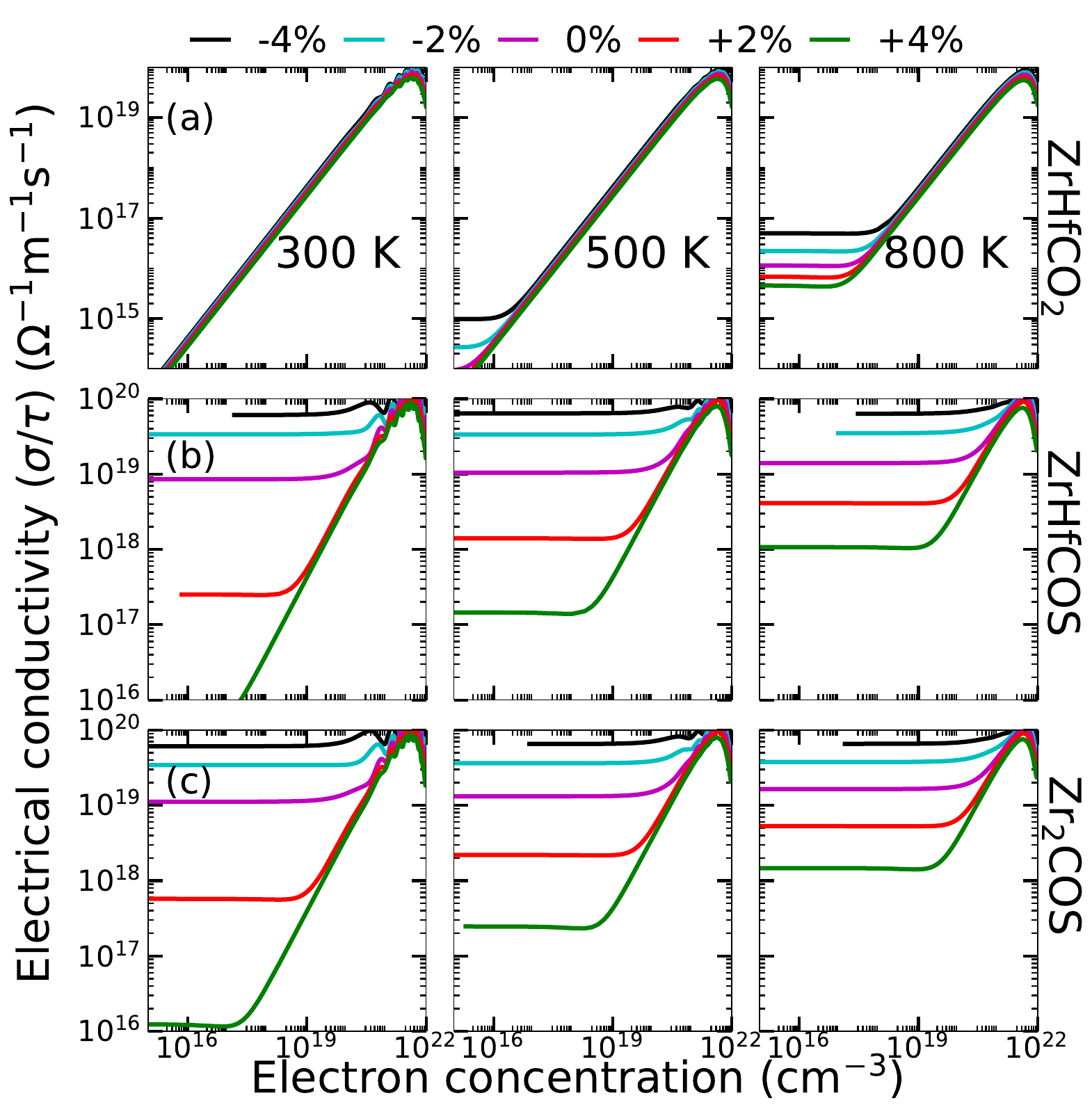}
    \caption{Electrical conductivity ($\sigma$/$\tau$) as a function of electron concentration (in cm$^{-3}$) for  n-type Janus MXenes ((a) ZrHfCO$_{2}$, (b) ZrHfCOS, and (c) Zr$_{2}$COS) at different temperatures. Results are shown for various $\epsilon$}
    \label{fig6}
\end{figure}

\begin{figure}
    \centering
    \includegraphics[scale=0.3]{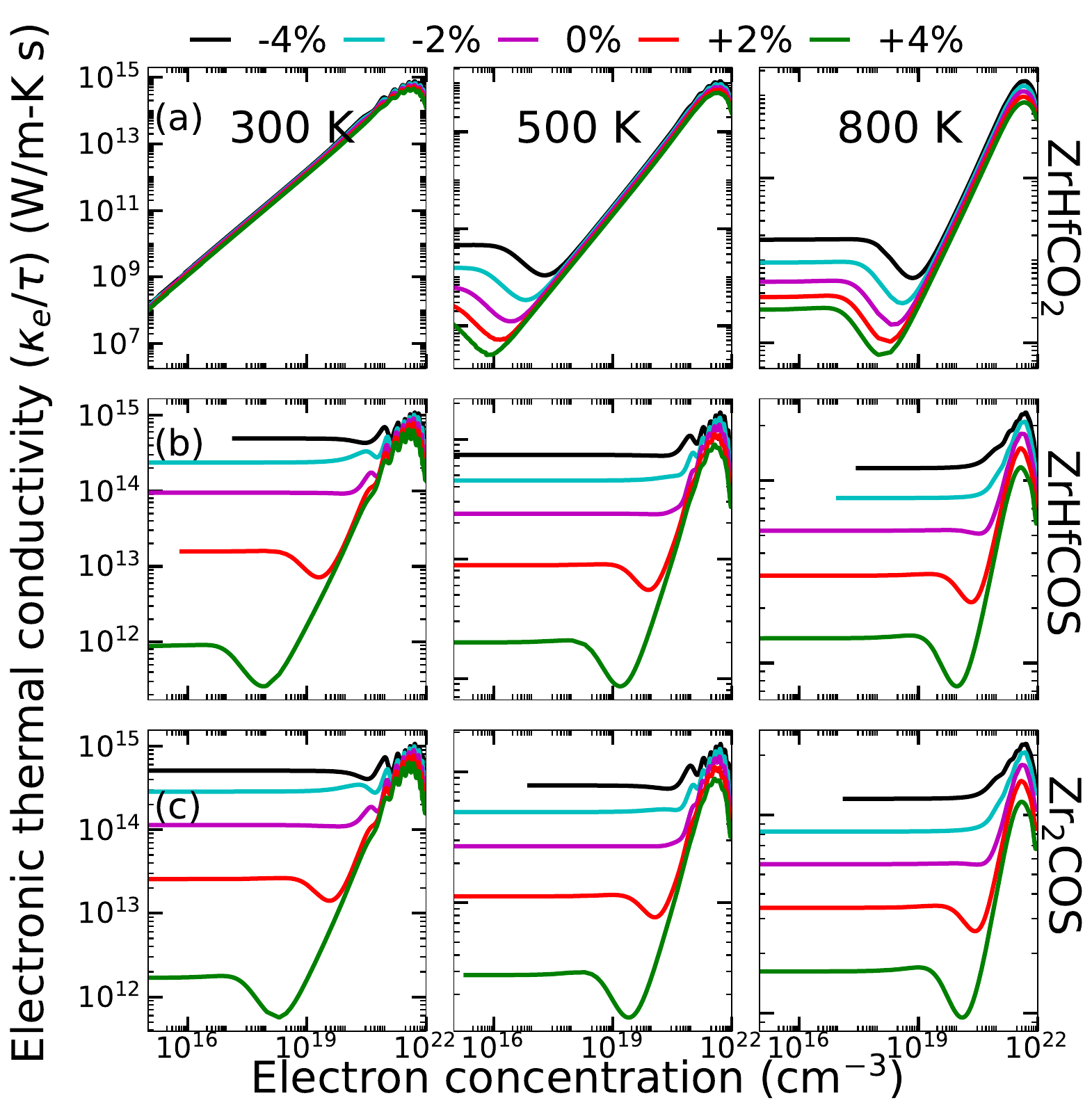}
    \caption{Electronic thermal conductivity ($\kappa_{e}$/$\tau$) as a function of electron concentration (in cm$^{-3}$) for n-type Janus MXenes ((a) ZrHfCO$_{2}$, (b) ZrHfCOS, and (c) Zr$_{2}$COS) at different temperatures.Results are shown for various $\epsilon$.}
    \label{fig7}
\end{figure}

Figure \ref{fig6},\ref{fig7} show electrical  conductivity($\sigma$) and electronic thermal conductivity ($\kappa_{e}$) of the three compounds, scaled by the relaxation time ($\tau$), as a function of n-type carrier concentration, strain and temperature. On comparing the $\sigma$/$\tau$ of the three Janus compounds, we find that $\sigma_{ZrHfCO_{2}} <\sigma_ {ZrHfCOS}\sim \sigma_{Zr_{2}COS}$. The behaviour is exactly opposite to that of $S$. In general, for high value of carrier concentration $n$, $\sigma$/$\tau$ increases linearly with $n$ which is consistent with standard $\sigma \propto n$ relation. However, a weak dependence of $\sigma$/$\tau$ on carrier concentration is observed for smaller $n$. The $\kappa_{e}$/$\tau$ follows the trends of $\sigma$/$\tau$, validating the Wiedemann-Franz law ($\kappa_{e}\propto\sigma$). Similar variations in $\sigma$/$\tau$ and $\kappa_{e}$/$\tau$ are found when the carrier is p-type (Figures S5 and S6, supplementary information). Thus, in all three Janus MXenes,  the biaxial strain is found to affect the electronic transport properties substantially. The conversion efficiency, nevertheless, is  constrained by the inverse dependence of $S$ and $\sigma$. The lattice thermal conductivity may, thus, be decisive.
\\
\section{Phonon dispersion and Lattice thermal conductivity}

The phonon dispersion relations and densities of states shown in Figure \ref{fig3} and S3, supplementary information, provide hints at possible effects of surface and strain engineering on lattice thermal conductivity $\kappa_{l}$. For all three Janus compounds, quadratic behaviour of ZA mode near $\Gamma$, hallmark of 2D systems, is observed. Phonon densities of states (Figure \ref{fig3}) show that the acoustic and first three optical modes are dominated by the vibrations from the transition metal atoms. C and the functional group elements (O and S) dominate the rest of the frequency spectra. As compared to ZrHfCO$_{2}$, the phonon modes shift towards  lower frequencies in ZrHfCOS and Zr$_{2}$COS. This is due to inclusion of S in the system replacing O, as S is heavier.  As a result, the lower optical modes get closer to the acoustic modes (an effect of surface engineering). Such overlapping acoustic-optical phonon modes promote three-phonon scattering and restrict heat transfer. Similar behaviour is observed when biaxial strain is applied (Figure S3, supplementary information). The phonon modes of all three systems shift to lower (higher) frequencies if the applied strain is tensile (compressive). Such shifts in phonon modes imply that the material's Debye temperatures ($\theta_{D}$) and group velocity reduce\cite{slack1973nonmetallic}. Consequently, lattice thermal conductivities ($\kappa_{l}$) are expected to decrease  with tensile strain. 
 
\begin{figure}
    \centering
    \includegraphics[scale=0.3]{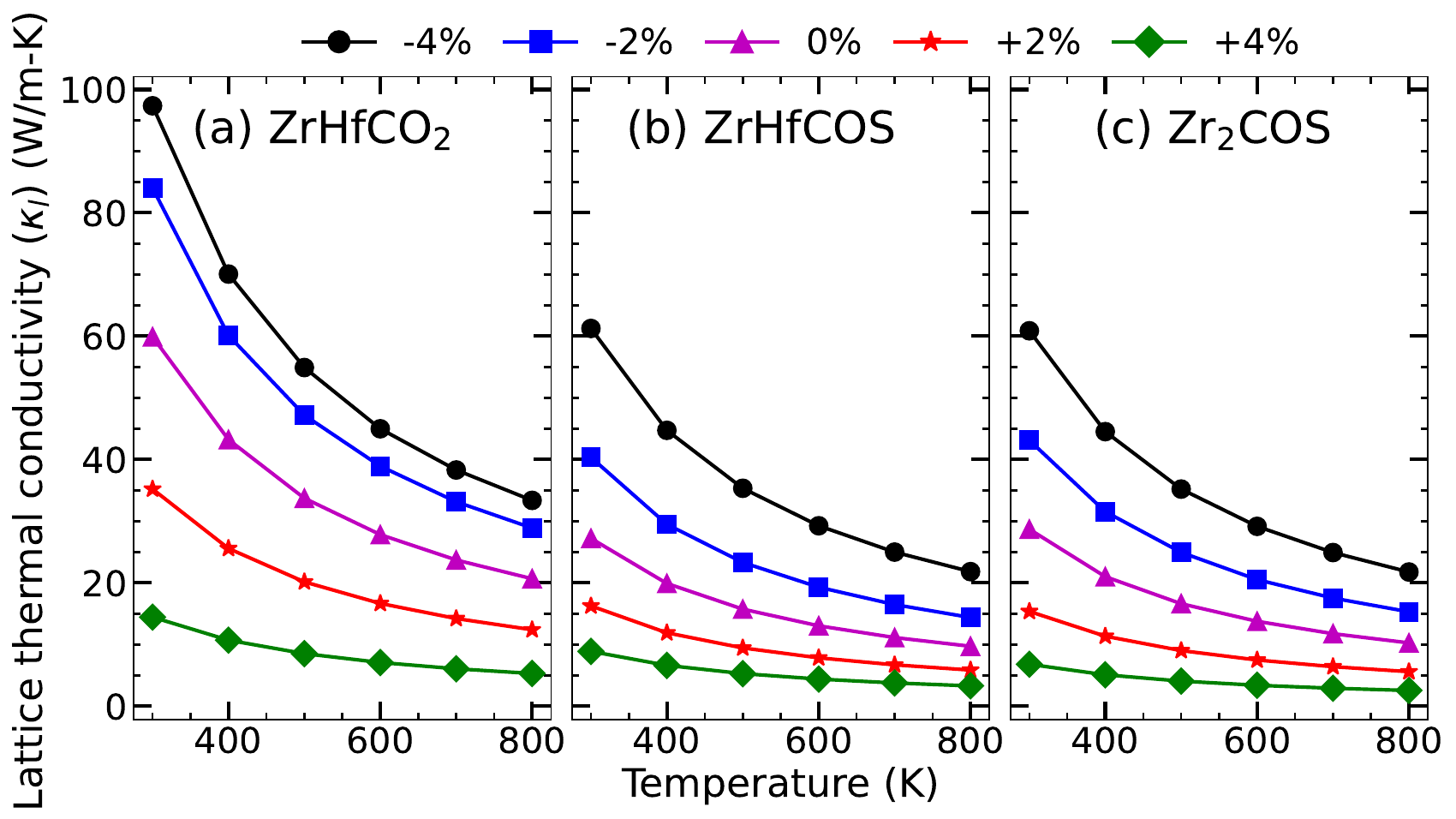}
    \caption{The lattice thermal conductivity as a function of temperature for the three compounds. Results are shown for different $\epsilon$.}
    \label{fig8}
\end{figure}

That this indeed is the case for the systems considered here is validated by the variations of $\kappa_{l}$ with strain. In Figure \ref{fig8} we present the variations in $\kappa_{l}$ as a function of temperature and strain for all three Janus MXenes. To substantiate the accuracy of our calculations, the convergence of $\kappa_{l}$ with respect to  $q$-grid size is shown in Figure S7 (supplementary information). For each case, the calculated $\kappa_{l}$ varies inversely with temperature. This suggests that intrinsic three-phonon (anharmonic) scattering dominates as temperature increases. We find that without any strain,  ZrHfCOS (ZrHfCO$_{2}$) has the lowest (largest) $\kappa_{l}$ of 27.19 W/m-K (59.88 W/m-K) at 300 K. This means that increasing surface asymmetry by heterogeneous passivation reduces $\kappa_{l}$ more (a reduction of about 55\% is observed in this case). Tensile biaxial strain in these compounds reduce $\kappa_{l}$ further. At 300 K, compared to the zero strain case,  $\kappa_{l}$ reduces by $\sim$ 76\% for ZrHfCO$_{2}$ and Zr$_{2}$COS , while it is 67\%,  for ZrHfCOS. The lowest $\kappa_{l}$ found for ZrHfCO$_{2}$, ZrHfCOS, and Zr$_{2}$COS at +4\% strain and 300 (800) K are 14.42 (5.28) W/m-K, 8.85 (3.28) W/m-K and 6.79 (2.53) W/m-K, respectively. These values  are significantly less than those of some well-studied TMDCs \cite{mose2,tmdc_kl}. 
\begin{figure}
    \centering
    \includegraphics[scale=0.25]{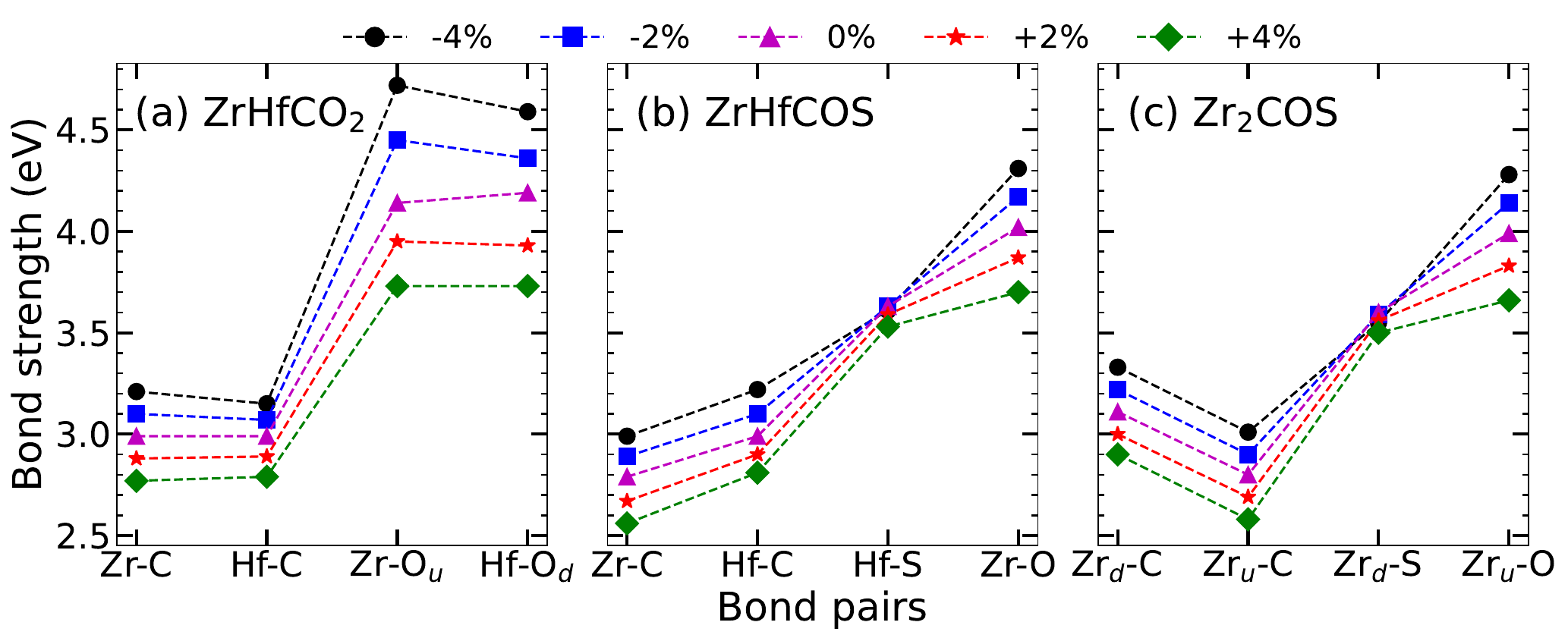}
    \caption{Bond strength (ICOHP (in eV)) for different atomic pairs of the three compounds. Results are shown for different $\epsilon$.}
    \label{fig9}
\end{figure}

In order to understand the reasons behind such drastic reductions of $\kappa_{l}$ with increasing asymmetry and tensile strain, we have performed analysis from various angles. First, we look into it from the perspective of structural parameters. To this end, we have done a qualitative analysis of the strengths of bonds between different atomic pairs as weaker (stronger) bond strengths imply larger (smaller) anharmonicity and lower (higher) $\kappa_{l}$ as a consequence. Significant variations in the bond lengths in these compounds, particularly with changes in the composition (Table \ref{tab1}) indicate dispersions in the bond strengths. Using the Crystal orbital Hamilton Population (COHP) method\cite{deringer2011crystal} as implemented in the LOBSTER package\cite{maintz2016lobster}, we have estimated the bond strengths (Integrated COHP or ICOHP). The results are shown in Figure \ref{fig9}. The results provide some insights regarding connections between bond strengths, strain and surface engineering to understand the trends in $\kappa_{l}$. In case of zero and compressive strains only, there are noticeable differences between the M-T bond strengths. The two compounds having S in one of the surfaces have weaker M-T bonds; the strengths of M-T bonds of these two compounds are comparable. The difference surely is due to the presence of longer bonds associated with S atoms. Also, irrespective of strain, the dispersions in bond strengths is more in the two compounds containing S. This, along with stronger bonds at zero and compressive strains in ZrHfCO$_{2}$ as compared to the other two compounds and subsequent large value of $\kappa_{l}$ indicate larger anharmonicity in S containing compounds. With increase in tensile strain, such differences reduce considerably. The reason lies in the relative changes of the M-T bond lengths. In ZrHfCO$_{2}$, the Hf-O bond length increases by 7\% as the strain changes from -4\% to +4\%. The corresponding bond in the other two compounds, that is Hf-S in ZrHfCOS and Zr$_{d}$-S in Zr$_{2}$COS change by only 1\% over the same changes in strain. As a result, the Hf-O bond strength reduces substantially with increase in strain in ZrHfCO$_{2}$ while there is hardly any change in the strengths of Hf-S and Zr$_{d}$-S bonds in ZrHfCOS and Zr$_{2}$COS, respectively.   
\begin{figure}
    \centering
    \includegraphics[scale=0.40]{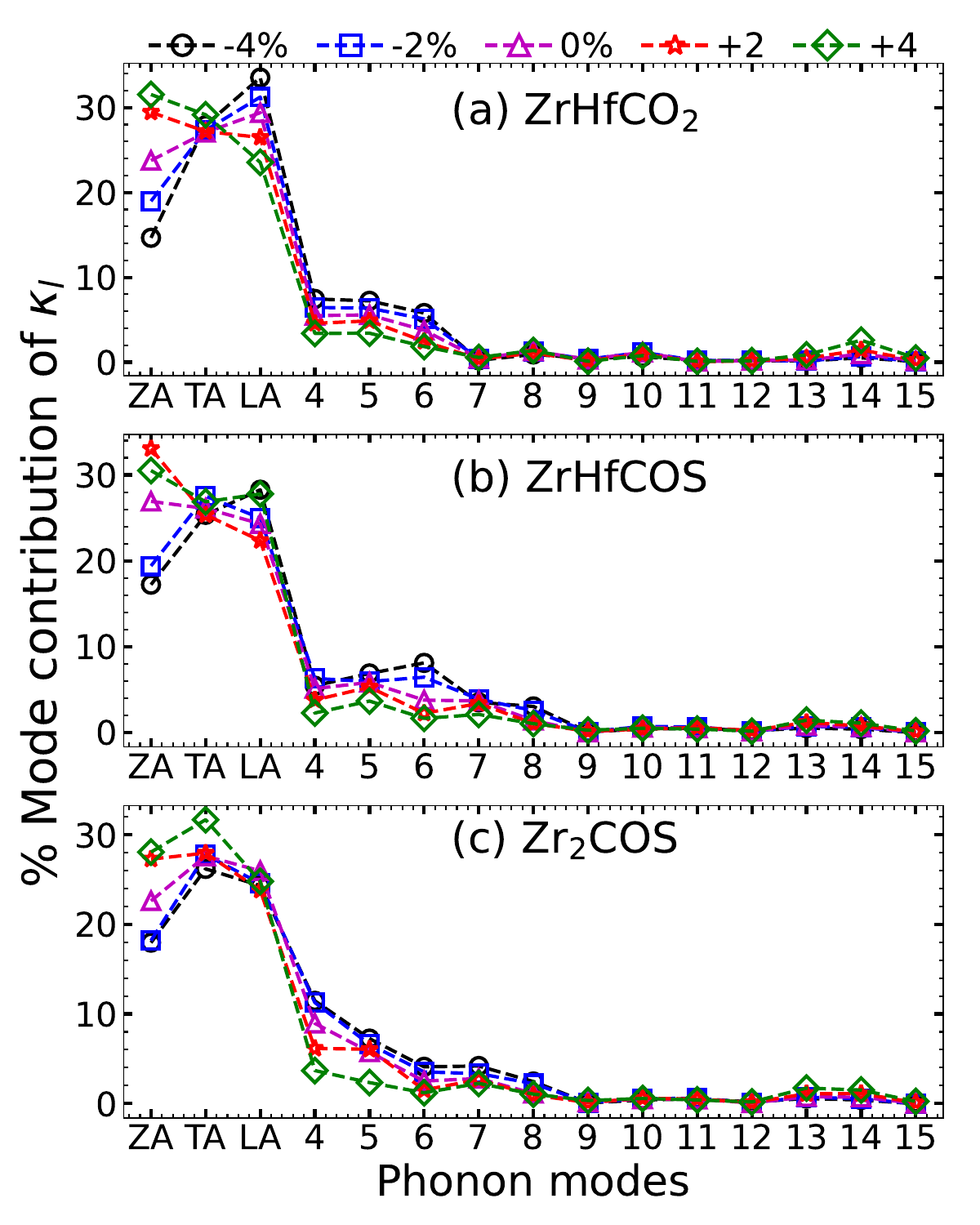}
    \caption{Phonon mode resolved contributions to lattice thermal conductivity (in \%) for the three compounds at  300 K. Results are shown for different $\epsilon$.}
    \label{fig10}
\end{figure}

Next we focus on the quantitative estimates of anharmonicity in the systems and the roles of surface and strain engineering. To this end, the first thing is identification of the largest contributors towards heat transport from among the phonon modes. We do this by calculating the percent contribution of each of the phonon modes to $\kappa_{l}$ (Figure \ref{fig10}). Across the compounds, irrespective of the strain, the major contributions are from the acoustic modes ($\sim$ 80\%) and the first three optical modes ($\sim$15\%). The  noticeable differences are (a) for compressive (tensile) strains, LA (ZA) mode is the major contributor among acoustic modes, and (b)  the contribution from the first three optical modes is more significant in case of compressive strains. Since these results suggest that frequencies lower than 6 THz are the primary contributors to $\kappa_{l}$, further analyses will be restricted in this frequency region only. Moreover, for the sake of qualitative understanding, cases with $\epsilon=-4,0,+4 \%$ only will be discussed.

\begin{figure}
    \centering
    \includegraphics[scale=0.2]{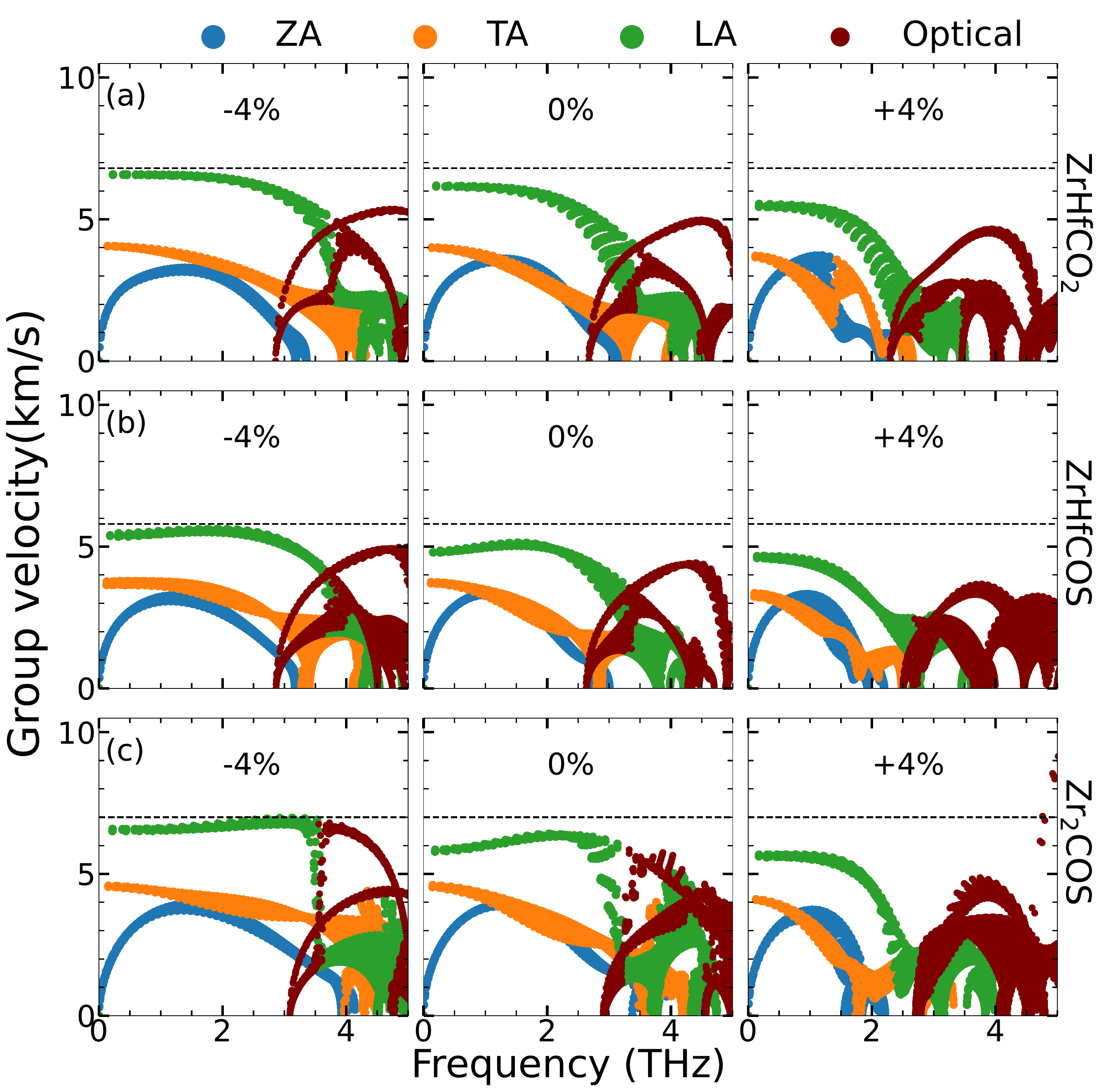}
    \caption{Phonon group velocity ($v_{g}$ (in km/s)) for the three compounds shown as function of phonon frequency. The value of $\epsilon$ in each case is given in the inset. The dashed line in each panel marks the highest $v_{g}$ for the parameters associated with the panel.}
    \label{fig11}
\end{figure}

As discussed earlier, the changes in the phonon dispersions with increasing strain point towards drop in phonon group velocities and consequently $\kappa_{l}$. The variations in group velocity $v_{g}$ at three values of $\epsilon$ are shown in Figure \ref{fig11}. We find that for each of the three compounds $v_{g}$ decreases (increases) with tensile (compressive) strain. This indeed explains the variations in $\kappa_{l}$ with $\epsilon$ for a given compound as $\kappa_{l}\propto v^{2}_{g}$ (Equation \ref{kl_eq}). Moreover, the quantitative comparison among $v_{g}$ of the three compounds at any $\epsilon$ including $\epsilon=0$ is consistent with the qualitative variation of $\kappa_{l}$ across compounds for a given bi-axial strain.  

\begin{figure}
    \centering
    \includegraphics[scale=0.25]{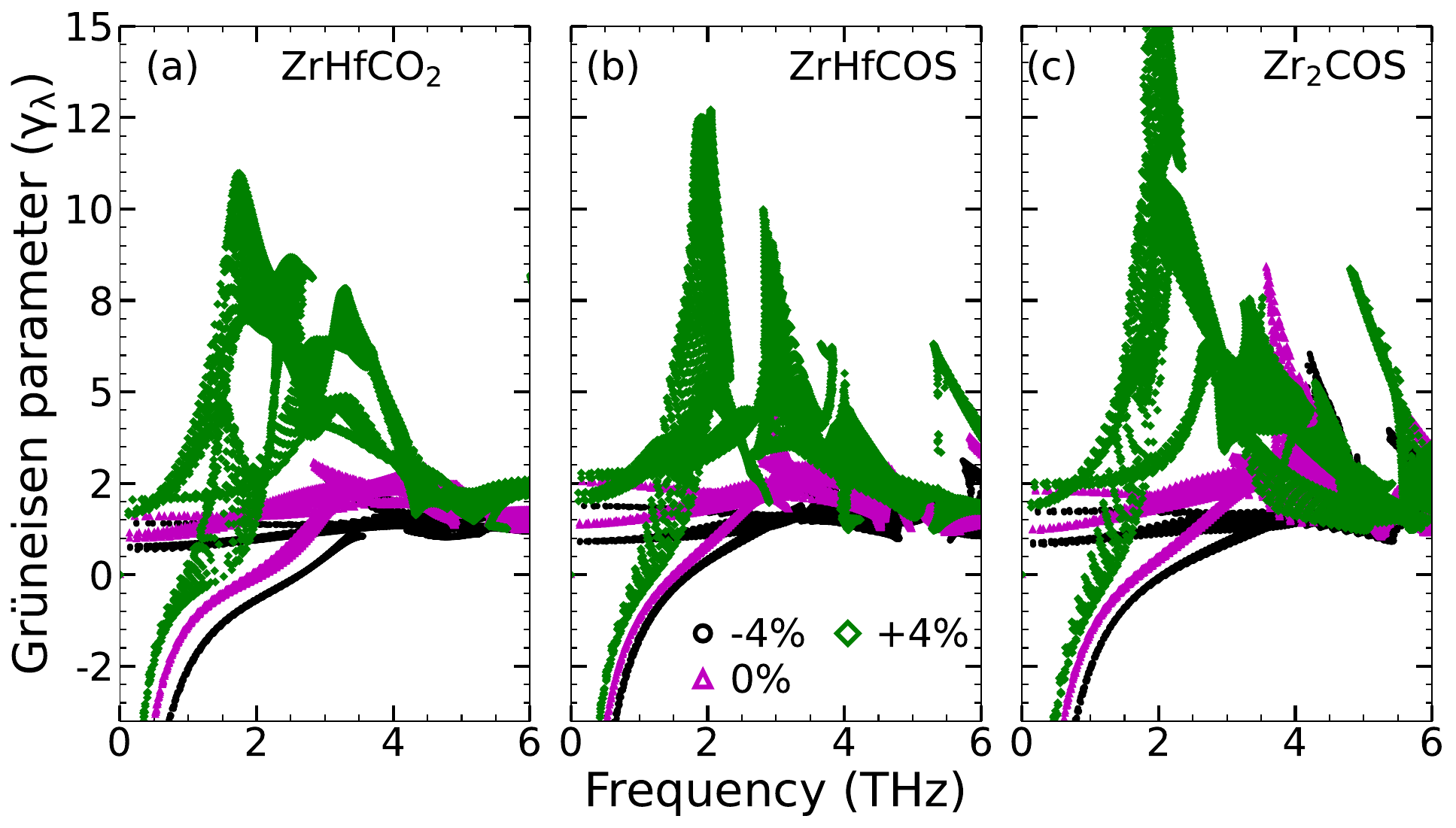}
    \caption{Gr\"uneisen parameter ($\Gamma_{\lambda}$) for three compounds at 300 K . Black, magenta and green indicate results for $\epsilon=-4,0,4$\%, respectively. }
    \label{fig12}
\end{figure}

Mode resolved Gr\"{u}neisen parameter ($\Gamma_{\lambda}$) is one of the parameters that quantify the degree of anharmonicity in a compound. In Figure \ref{fig12}, we present  results of $\Gamma_{\lambda}$ as function of phonon frequencies, calculated using anharmonic IFCs\cite{gruneisen}. At zero and 4\% tensile strain, Zr$_{2}$COS has largest $\Gamma_{\lambda}$ followed by ZrHfCOS and ZrHfCO$_{2}$. This trend is consistent with the trend in $\kappa_{l}$ at 300 K. For a compressive strain of -4\%, larger and near equal contributions towards $\Gamma_{\lambda}$ between 2 and 4 THz for ZrHfCOS and Zr$_{2}$COS in comparison to ZrHfCO$_{2}$ too explains the trends in $\kappa_{l}$ among the compounds and strain values. 

\begin{figure}
    \centering
    \includegraphics[scale=0.25]{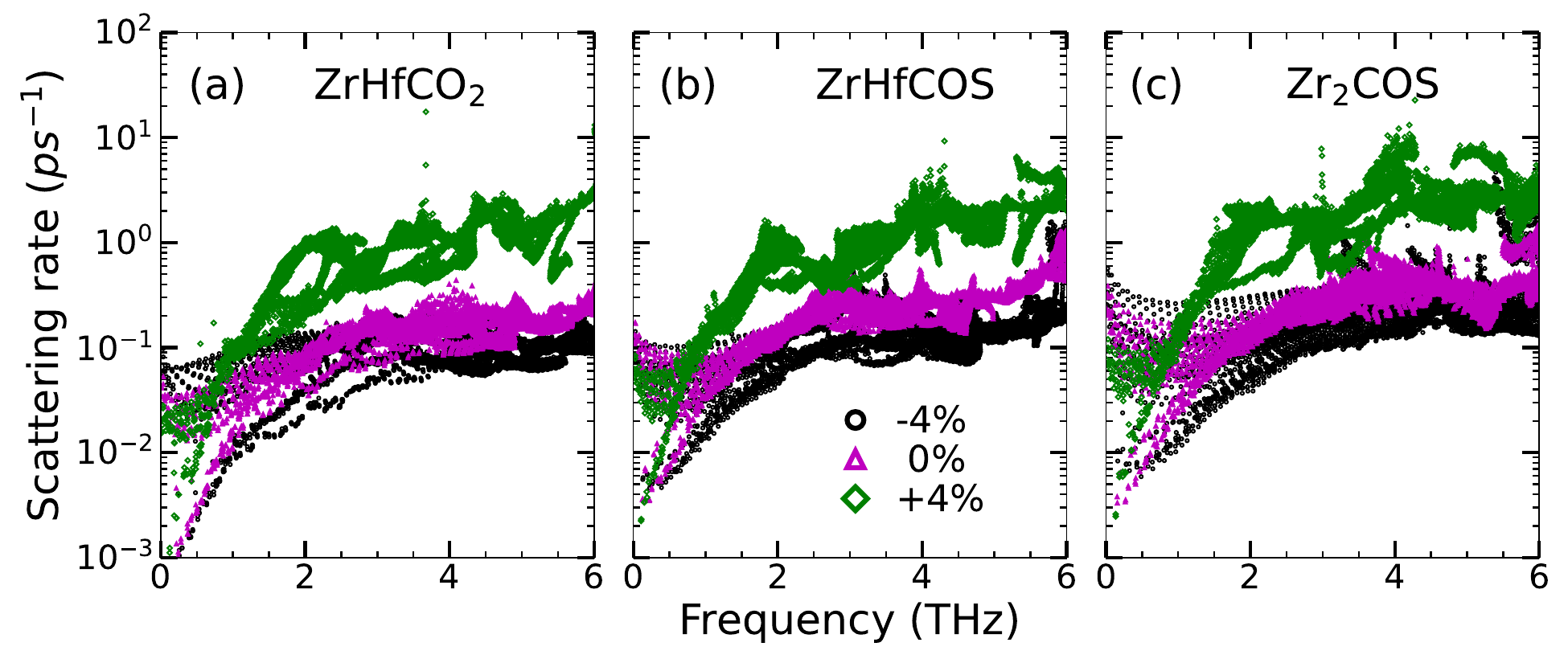}
    \caption{Anharmonic scattering rates for three compounds at 300 K . Black, magenta and green indicate results for $\epsilon=-4,0,4$\%, respectively.}
    \label{fig13}
\end{figure}

In order to understand the trends in the strengths of anharmonicity as seen in the variations of $\Gamma_{\lambda}$ in terms of phonon phonon scattering, we next look at the anharmonic scattering rates. Figure \ref{fig13} shows the calculated anharmonic scattering rates for the relevant frequency region. The higher scattering rates for S containing Janus in comparison with ZrHfCO$_{2}$ at any $\epsilon$ including $\epsilon=0$ explains the trends in $\kappa_{l}$ with composition and strain. To understand the reason behind higher scattering rates in S containing Janus, we have evaluated the weighted scattering phase space (WP$_{3}$) accessible to the compounds (Figure S8, supplementary information). Results are shown for three values of $\epsilon$. For each compound, we find that WP$_{3}^{+4\%}$ $>$ WP$_{3}^{0\%}$ $>$ WP$_{3}^{-4\%}$. This explains the trends in the scattering rates and thus that of $\kappa_{l}$. However, the information obtained from WP$_{3}$ do not have the clarity to explain higher anharmonic scattering rates in Janus containing S. We, therefore, conclude that the higher anharmonic scattering rates in Janus containing S in comparison with ZrHfCO$_{2}$ for compressive and zero strains can be understood from the comparative rates of scattering  between 1-4 THz. The greater proximity of the acoustic and optical branches due to overall downward shift of phonon frequencies brought about by the replacement of O with heavier S leads to higher scattering rates in these two Janus compounds. This is the important contribution of surface engineering.    

\section{Figure of merit ($ZT$)}
\begin{figure}
	\centering
	\includegraphics[scale=0.21]{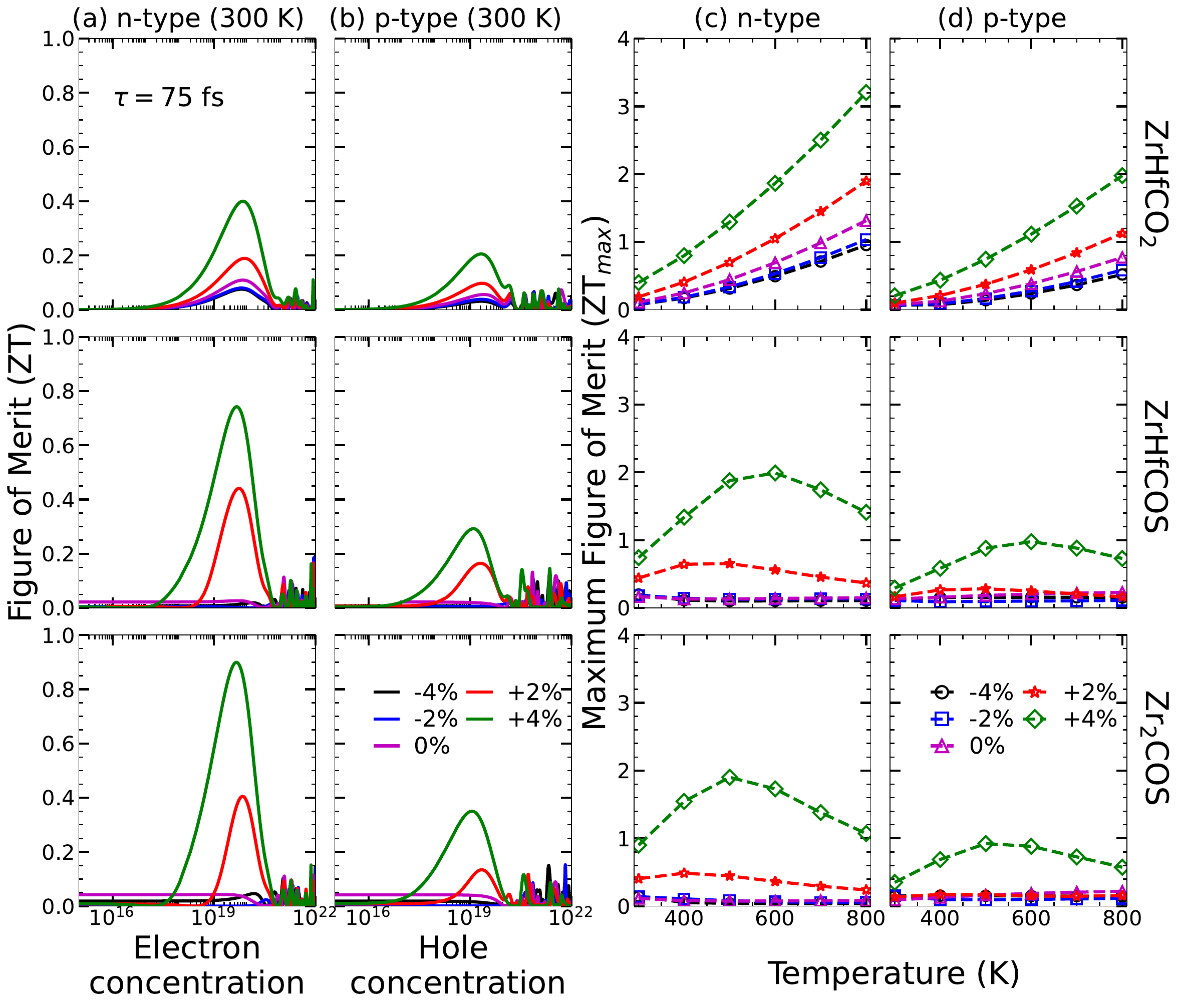}
	\caption{Figure of merit as a function of carrier concentration at 300 K (column a,b) and maximum figure of merit as a function of temperature (column c,d) for the three Janus MXenes. Results are shown for various $\epsilon$. The calculations are done with carrier relaxation time $\tau=75fs$.}
	\label{fig14}
\end{figure}

Apart from temperature, $ZT$ depends on the carrier concentration $n$ and relaxation time $\tau$ since the transport parameters are calculated under RBA and CRTA. In Figure \ref{fig14} (a)-(b), we show the variation of  $ZT$ as a function of carrier concentrations for different strains, at 300 K. For n-type (p-type) doping of compounds, the variations of maximum $ZT$ with strain and temperature are shown in Figure \ref{fig14}(c) (Figure \ref{fig14}(d)). The results presented are for $\tau$=75 fs. We have chosen this particular value of $\tau$ as values close to it has been used successfully for various 2D thermoelectrics \cite{zt_tau1} including MXenes \cite{tanusri,jing2019superior}. $ZT$ as a function of carrier concentration, strain (at 300 K) and maximum $ZT$ as a function of temperature and strain for each compound calculated for different $\tau$ are shown in Figures S9-12, supplementary information. The results suggest that irrespective of value of $\tau$, maximum $ZT$ is obtained around the carrier concentration of 10$^{19}$-10$^{20}$ cm$^{-3}$. This, too, agrees with the results of previous works. Irrespective of the systems and the type of carrier, the maximum $ZT$ increases as one moves from the region of compressive to that of the tensile strain. However, the $ZT$ values for n-doped systems are higher in all cases. The increase (decrease) of maximum $ZT$ with tensile (compressive) strain is due to reduction (elevation) in $\kappa_{l}$ while larger (smaller) $ZT$ for n-type (p-type) doping is due to the trends in the electronic transport parameters. The increase of maximum $ZT$ with $\tau$ for all systems, irrespective of $\epsilon$ is an artefact of increase in the electronic transport coefficients with $\tau$. Finally, we find substantial impact of surface and strain engineering in the variation of maximum $ZT$ with temperature. Although the qualitative effects of strain in this case is identical for all three systems, the values and the qualitative variations differ across the systems. For the maximum tensile strain considered, a maximum $ZT$ of 3.2 is obtained in ZrHfCO$_{2}$ at 800 K while for ZrHfCOS (Zr$_{2}$COS) it is $\sim 2$ at 600 K (500 K). This difference is mostly due to the variations in the electronic transport parameters, notably the Seebeck coefficient $S$. Since with increase in temperature, the differences of $\kappa_{l}$ among the three compounds for the maximum tensile strain reduce considerably in comparison with that at 300 K, the differences in maximum $ZT$ is purely due to electronic transport. Nevertheless, the maximum $ZT$ for all three compounds, obtained at a tensile strain of 4\% are close to those obtained in other surface-engineered MXenes \cite{tanusri,murari2024symmetry}. Moreover, such high $ZT$ values are desirable to obtain from the thermoelectric device, a Carnot efficiency at par with a Carnot refrigerator.

\section{Conclusions}
Using first-principles DFT and semi-classical Boltzmann transport theory, the microscopic understanding of the effects of strain and surface engineering in improvement of thermoelectric efficiency in 2D MXenes has been done in this work through a rigorous analysis of their structural, electronic and dynamical properties. The surface engineering is done by creating different types of asymmetric surfaces in M$_{2}$C MXenes. We find that while the strain is the primary factor affecting the electronic transport parameters like Seebeck coefficient $S$, electrical conductivity $\sigma$ and electronic thermal conductivity $\kappa_{e}$, both composition (surface manipulation) and strain substantially affect the lattice thermal conductivity $\kappa_{l}$. As a consequence, we obtain excellent thermoelectric figure of merit $ZT$ for all three compounds within a temperature window of 500 K. The tensile strain on one hand modifies the electronic band gap and the electronic structures of the compounds, thus improving the electronic parameters, while on the other hand it increases anharmonicity through enhancement of phonon-phonon scattering. The interplay of the electronic and lattice transport parameters in the three compounds that differ in their compositions is understood to be the reason behind higher maximum $ZT$ obtained in ZrHfCO$_{2}$, inspite of undergoing the least engineering of its surfaces. This work establishes that application of moderate tensile strain on Janus systems, obtained from manipulation of surfaces through the functional groups attached to the transition metal components in 2D materials like MXenes, can be an efficient way to obtain materials with potentially high thermoelectric conversion efficiencies.     

\section{Acknowledgement}
The authors gratefully acknowledge the Department of Science and Technology, India, for the computational facilities under Grant No. SR/FST/P-II/020/2009 and IIT Guwahati for the PARAM supercomputing facility.


%
\end{document}